\documentclass[aps,prc,twocolumn,floatfix,nofootinbib,preprintnumbers,superscriptaddress,amsmath,amssymb]{revtex4}

\usepackage{graphicx}
\usepackage{dcolumn}
\usepackage{bm}
\usepackage{hyperref}
\usepackage[mathlines]{lineno}

\begin{document}

\title{ Continuum damping effects in nuclear collisions associated with twisted boundary conditions  }

\author{C.Q. He}
\affiliation{State Key Laboratory of Nuclear
Physics and Technology, School of Physics, Peking University,  Beijing 100871, China}

\author{J.C. Pei}
\email{peij@pku.edu.cn}
\affiliation{State Key Laboratory of Nuclear
Physics and Technology, School of Physics, Peking University,  Beijing 100871, China}

\author{Yu Qiang}
\affiliation{State Key Laboratory of Nuclear
Physics and Technology, School of Physics, Peking University,  Beijing 100871, China}

\author{Na Fei}
\affiliation{State Key Laboratory of Nuclear
Physics and Technology, School of Physics, Peking University,  Beijing 100871, China}

\begin{abstract}

The time-dependent Skyrme Hartree-Fock calculations have been performed to study $^{24}$Mg +$^{24}$Mg collisions.
The twisted boundary conditions, which can avoid finite box-size effects of the employed 3D coordinate space,  have been implemented.
The prolate deformed $^{24}$Mg has been set to different orientations to study vibrations and rotations of the compound nucleus $^{48}$Cr.
Our time evolution results show continuum damping effects associated with the twist-averaged boundary condition play a persistent role after the fusion stage.
In particular, a rotational damping in continuum is presented in calculations of both twist-averaged and absorbing boundary conditions, in which damping widths can be clearly extracted.
It is unusual that the rotating compound nucleus in continuum evolves towards spherical but still has a considerable angular momentum.

\end{abstract}

\maketitle


\emph{Introduction.}---
The real-time nuclear dynamics such as collective responses, large amplitude collisions and  fissions have been studied extensively to probe effective interactions, many-body correlations and transport properties~\cite{negele,tddft,tdhf}.
The basic theoretical framework for quantum many-body dynamics is the time-dependent Schr\"{o}dinger equation with various approximations.
In this respect, the microscopic time-dependent Hartree-Fock (TDHF) (or time-dependent density functional theory) was very successful for studies of nuclear dynamics, particularly the large amplitude dynamics~\cite{negele,tddft,tdhf,umar,guo,scamps,paul}.
The improved time-dependent-Hartree-Fock-Bogoliubov calculations have also been developed for superfluid systems, relying on tremendous computing capabilities~\cite{tdhfb1,tdhfb2}.
Besides, the quantum molecular dynamics calculations have been widely used for heavy ion collisions at higher energies with two-body dissipations~\cite{qmd}.
In the small amplitude limit of collective motions, TDHF results can match the random phase approximation (RPA), or linear resonance theory~\cite{tddft,umar05}.
For nuclear collisions involving considerable excitation energies,  the TDHF without pairing is a reasonable approximation.
For too high excitation energies, the TDHF is not applicable when two-body collisions become prominent.

TDHF calculations are usually performed in 3D coordinate spaces with periodic boundary conditions~\cite{sky3d}. Due to the limits of computational resources,
the calculations suffer from a systematic error related to finite box-sizes that were employed.  The finite box sizes lead to tinny wave reflections at boundaries or
wave interferences between periodic images~\cite{abc1}.
Furthermore, the continuum can not be accurately discretized within small coordinate spaces~\cite{abc2}.
This is not a serious problem for descriptions of bulk properties but it could be not negligible after long-time evolutions.
The important role of continuum in nuclear reactions has also been demonstrated by the widely adopted continuum-discretized coupled-channels calculations~\cite{cdcc}.
For highly excited compound nuclei produced by fusion reactions, the surface pressures are equilibrized by thermalized continuum gases which allow for particle evaporations~\cite{zhu}.
In weakly bound nuclei, the accurate treatment of continuum couplings is important for halo structures and associated dynamics~\cite{matsuo,zhang,sun}.

To avoid the box-size dependence in the treatment of continuum, the twisted averaged boundary condition (TABC) has been applied to TDHF calculations of giant resonances~\cite{schuer}.
The twisted boundary condition (TBC) is a generalized periodic boundary condition (PBC) for Bloch waves with non-zero twisted angles.
In condensed-matter physics, TABC results by averaging twisted angles can significantly cancel finite box-size effects~\cite{lin}.
Partial TBC has also been found to be useful in Lattice QCD calculations of few body systems~\cite{luu}.
In nuclear physics, the quasiparticle RPA (QRPA) with outgoing boundary conditions has been realized for spherical nuclei~\cite{matsuo} but it is extremely difficult for deformed nuclei.
It was known that QRPA calculations with not-well discretized continuum can cause false resonance peaks~\cite{abc2,wang}.
An effective way to smooth resonances of deformed nuclei is to use the absorbing boundary condition (ABC)~\cite{abc1,abc2}.
It has been demonstrated that the TABC and ABC behave similarly in damping effects to smooth giant resonances~\cite{schuer}.
The TDHF calculations can in principle take into account Landau damping, escaping damping and and  the damping due to complex configuration couplings~\cite{tdhf}.
The continuum treatment is essential in all of these damping mechanisms.
The ABC calculation has to adjust the imaginary potential to absorb waves exactly at boundaries, and this is tedious.
On the other hand, TABC can be easily implemented for complex systems.
The TABC is very successful in studying giant resonances which are considered as small amplitude collective motions.
Therefore, it is desirable to explore the influences of TABC calculations of large amplitude nuclear collisions with time evolutions.

In this work, we intend to study the $^{24}$Mg +$^{24}$Mg collisions by TDHF calculations with TABC, PBC and ABC boundary conditions.
The compound nucleus $^{48}$Cr can have hyperdeformed states at high spins from cranking calculations~\cite{high-spin}, indicating multi-$\alpha$ clustering structures. Such $\alpha$-conjugate  compound nuclei are expected
to be favorable for searching collective molecular motions. Indeed, several experiments have been performed for $^{24}$Mg +$^{24}$Mg collisions and
narrow resonances in inelastic cross sections have been reported~\cite{wuossma,nitto,zurmuhle}.  However, some experiments didn't find resonance structures in $^{24}$Mg +$^{24}$Mg fusion cross sections~\cite{Jachcinski}.  Note that the full picture of clustering structures in compound nuclei should take into account the dynamical nonlocalized clustering~\cite{zhou}.
In the precompound nuclei, the time-dependent nucleon localization indicates that clustering vibrations are important in the initial stage of fusion~\cite{schuer3}.
The prolate $^{24}$Mg allows studies of collisions with different orientations. To characterize different collision reactions,
the time evolutions of total kinetic energies and deformations have been studied using the Fourier transformation.

\emph{Method.}---
We utilize the 3D Skyrme-TDHF solver Sky3D~\cite{sky3d}, which solves the self-consistent HF equation and the  TDHF equation.
Calculations are performed in the 3D uniform coordinate space, and there is no symmetry restrictions on the wavefunctions.
The full Skyrme energy functional adopts the SV-bas~\cite{svbas} force, in which the spin-orbit and time-odd terms have been included.

The grid spacing is set to be 1 fm and the time step of dynamical evolution takes 0.2 fm/c.
In Sky3D, the time propagator is evaluated by the Taylor series expansion up to the sixth order.
Computations with these settings have been demonstrated to be good enough for descriptions of the essence of dynamical properties~\cite{dai}.
The static calculations of $^{24}$Mg+$^{24}$Mg are firstly carried out to obtain the ground-state wave functions, which are inputs for time-evolution calculations.
The 3D box sizes (along x, y, z-axis) in static and dynamical calculations are taken as 24$\times$24$\times$24 fm and 48$\times$24$\times$48 fm, respectively. Note that the static wave functions can be transformed into larger coordinate spaces
by using the Fourier and inverse Fourier transformations.
The energies and density distributions as a function of time are the main outputs.

\emph{Boundary conditions.}---
PBC is a natural choice for plane wave representation and is efficient for computations in the uniform 3D grids~\cite{sky3d}.
TBC is a generalized Bloch boundary condition as written as~\cite{schuer},
\begin{equation}
   \psi(\textbf{r}+\mathbf{n}L)=e^{i\mathbf{\theta} \cdot \mathbf{n}} \psi(\textbf{r}) \textrm{,}
   \label{eq1}
\end{equation}
where $\mathbf{r}$ denotes the 3D coordinates, $L$ denotes the box size, and $\mathbf{n}$ is the unit vector in 3D Cartesian coordinates.
The twisted angle $\theta$ changes from zero to $\pi$.  Eq.$\ref{eq1}$ can go back to PBC when the twisted angle $\theta$ is zero.
The single-particle HF equation can be written as,
\begin{equation}
\hat{{h}}_{\theta}\psi_{\alpha \theta}(\mathbf{r})=\epsilon_{\alpha\theta}\psi_{\alpha \theta}(\mathbf{r})
\end{equation}
 where $\alpha$ is the discrete label of the single-particle wave functions.

In the TABC method, the expectation value of an observable \^{O} can be obtained by averaging over the twist angles~\cite{schuer},
\begin{equation}\label{2}
  \langle \hat{O}(t) \rangle=\frac{1}{8\pi^3} \iiint^{2\pi}_0 d^3\theta\langle \Psi_\theta(t)|\hat{O}|\Psi_\theta(t) \rangle
\end{equation}
where $\Psi_{\theta}(t)$ is the HF Slater wave function at time $t$.
The twisted angle is discretized in practical calculations.
In this work, the 3D integration over $\theta$ is performed using a four-point Gauss-Legendre quadrature between 0 and $\pi$.
This means that total 64 TDHF calculations are carried out for each case.
The momentum $\mathbf{\textit{k}}$ is modified accordingly as,
\begin{equation}
k_{i,m}= \frac{2\pi m+\theta_i}{L_i}, \hspace{4pt} m=0,\pm1, \pm2,  \cdots, \pm m_{max}.
\end{equation}
In principle, we can recover a continuous spectrum of $\bf{ \textit{k}}$ with varying twisted angles.
 Calculations
with different twist angles will give rise to different finite-volume corrections~\cite{luu}.
It has been demonstrated that
averaging results over $\theta$ can significantly cancel finite-volume effects~\cite{lin,schuer,schuer2}.

For comparison, we also implemented ABC using the mask function method~\cite{abc1,schuer}. It was known that the mask function
is effective as the imaginary absorbing potential~\cite{abc1}. The mask function applies to wavefunctions and is given as
${\rm cos}(\frac{\pi}{2}\frac{r-L/2+l_{abs}}{l_{abs}})^p$ for $L/2-l_{abs}$$<r$$\leqslant$$ L/2$.
In ABC calculations, 
the box sizes are taken as 56$\times$56$\times$56 fm.
In this case, the absorbing thickness $l_{abs}$ is 12 fm and $p$ is taken as 0.04. These values are dependent on
the specific box sizes and time steps. 

\begin{figure}[t]
  \includegraphics[width=0.56\textwidth]{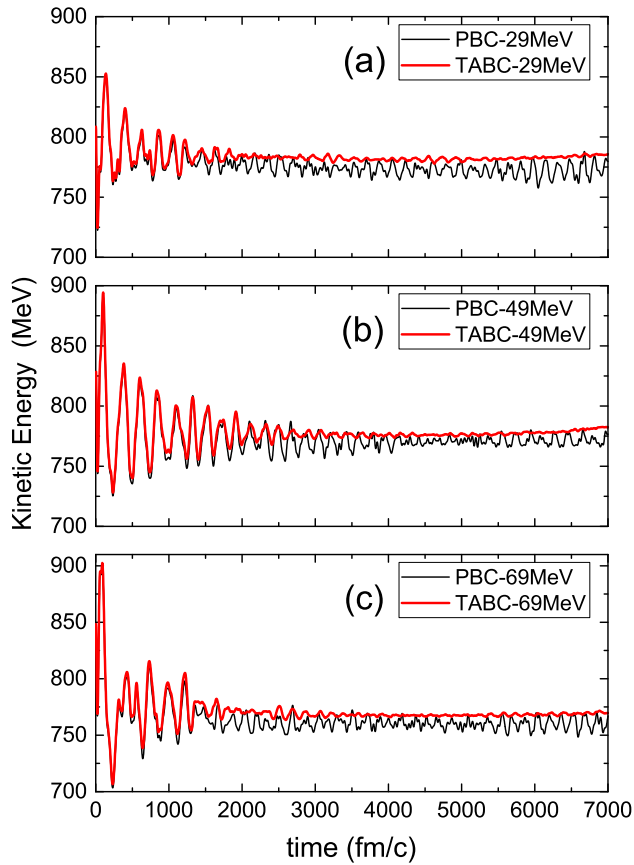}\\
  \caption{(Color online) TDHF calculated time evolution of kinetic energies for the head-to-head $^{24}$Mg+$^{24}$Mg collisions with PBC and TABC.
  (a)$E_{c.m.}$=29 MeV;(b)$E_{c.m.}$=49 MeV;(c)$E_{c.m.}$=69 MeV.}
  \label{fig1}
\end{figure}

\begin{figure}[t]
  \includegraphics[width=0.5\textwidth]{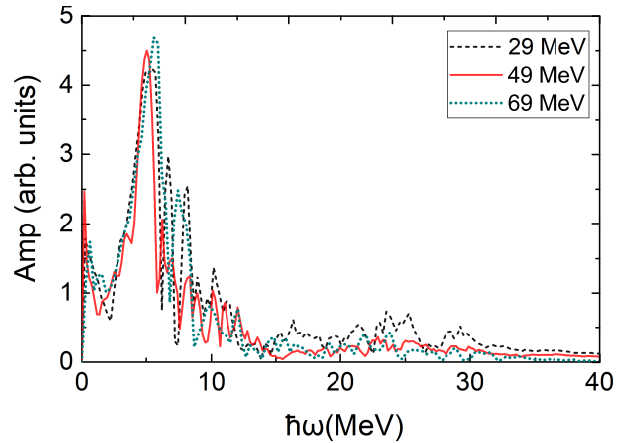}\\
  \caption{(Color online) Fourier analysis of the evolution of kinetic energies in TABC calculations for $E_{c.m.}$=29, 49, 69 MeV, corresponding to Fig.\ref{fig1}.}
  \label{fig2}
\end{figure}

\begin{figure}[t]
  \includegraphics[width=0.49\textwidth]{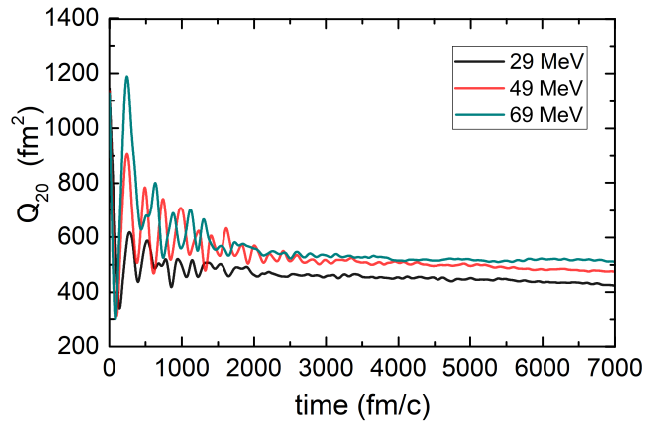}\\
  \caption{(Color online) Evolution of Quadrupole moments from TDHF calculations with TABC at $E_{c.m.}$=29, 49, 69 MeV, corresponding to Fig.\ref{fig1}. }
  \label{fig3}
\end{figure}

\emph{Results.}---	
We have performed 3D TDHF calculations using the Sky3D solver for $^{24}$Mg+$^{24}$Mg collisions.
The ground state of $^{24}$Mg has a large prolate deformation of $\beta_2$=0.49 (axis ratio is 1.7:1) in our Skyrme Hartree-Fock+BCS calculations.
The dimensionless quadrupole deformation is defined as $\beta_2$=$\frac{4\pi}{3AR_0^2}<r^2Y_{20}>$ ~\cite{bender}. 
The reaction threshold energy is -14.93 MeV since the binding energies of $^{24}$Mg and $^{48}$Cr are 198.26 MeV and 411.45 MeV~\cite{wangm}, respectively.
The fusion barriers in this case are from 22 MeV (head to head) to 24 MeV (side to side) depending on the collision orientation.

\begin{figure}[t]
  \includegraphics[width=0.58\textwidth]{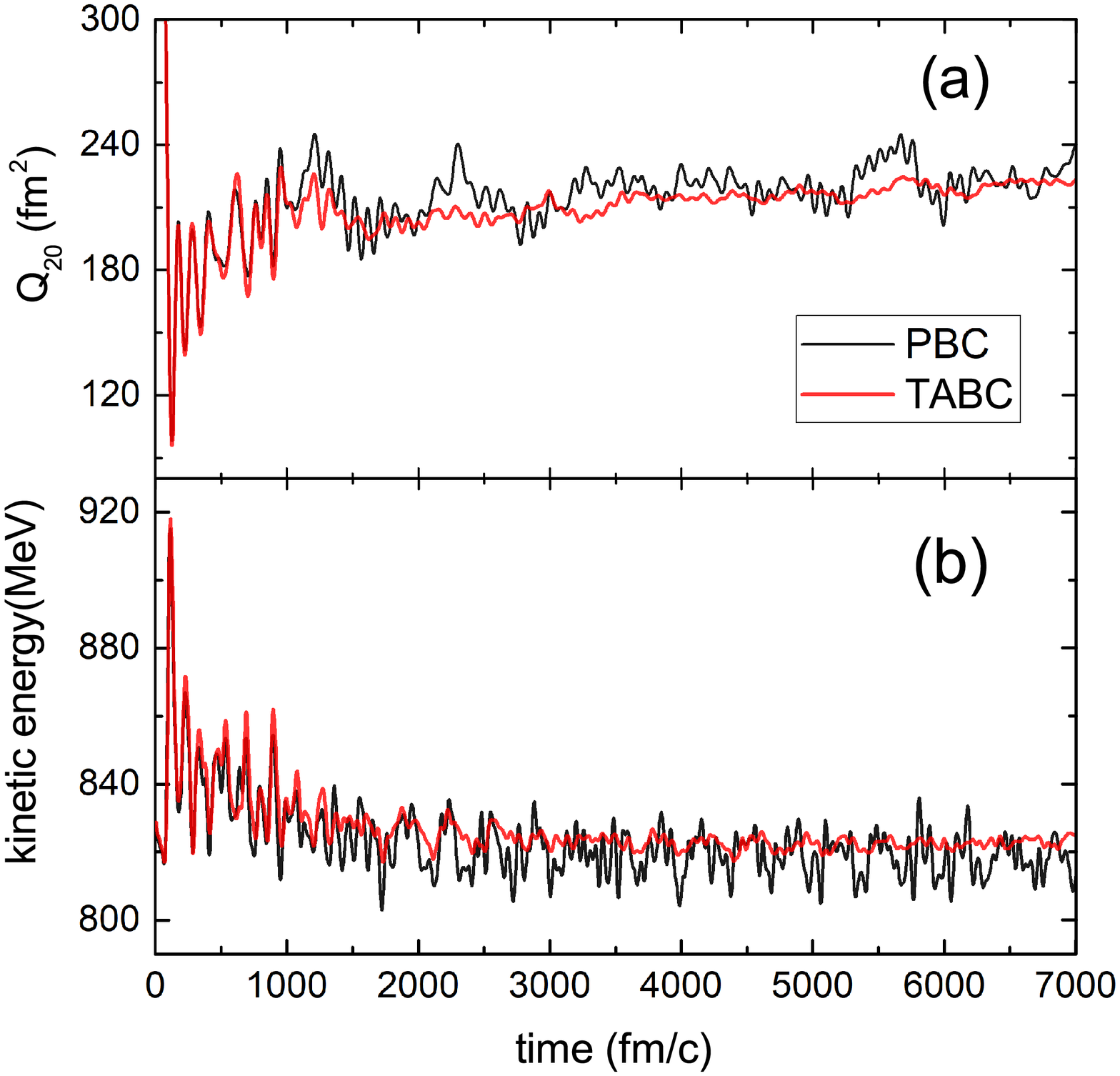}\\
  \caption{(Color online) The evolution of the quadrupole deformation and the kinetic energy of the side-to-side $^{24}$Mg+ $^{24}$Mg collision at  $E_{c.m}$=49 MeV.  }
  \label{fig4}
\end{figure}

\begin{figure}[t]
  \includegraphics[width=0.45\textwidth]{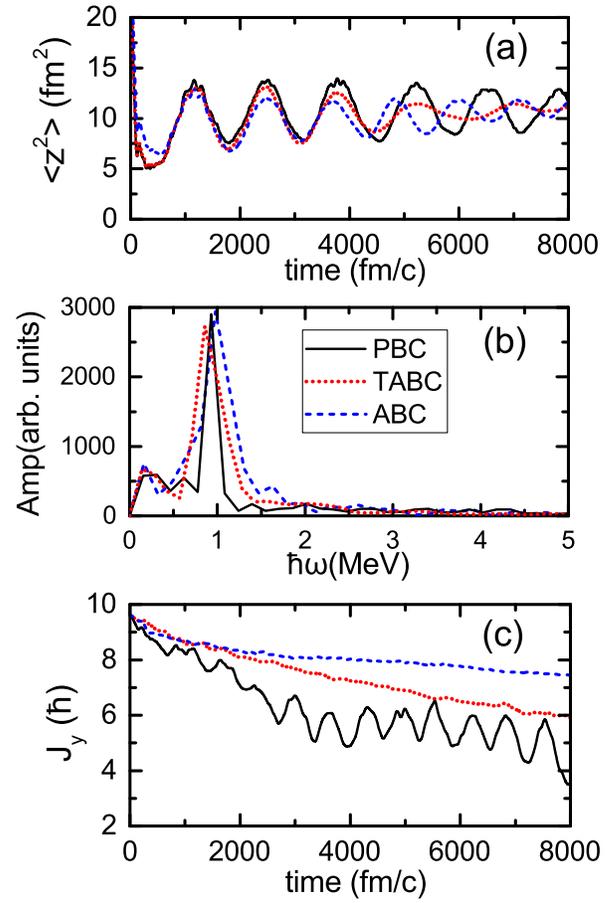}\\
  \caption{(Color online) The evolution results of the $^{24}$Mg+ $^{24}$Mg side-to-side collision with an impact parameter of 2 fm and $E_{c.m}$=40 MeV.
    The upper panel shows the time evolution of mean square value of ${<z^2>}$; the middle panel shows the Fourier analysis of the evolution results of the upper plane.
    The lower panel shows the time evolution of the expection value of the angular momentum $<$$J_y$$>$.
     Red line,  black line and blue line denote TABC, PBC and ABC results respectively.  }
  \label{fig5}
\end{figure}

\begin{figure}[t]
  \includegraphics[width=0.48\textwidth]{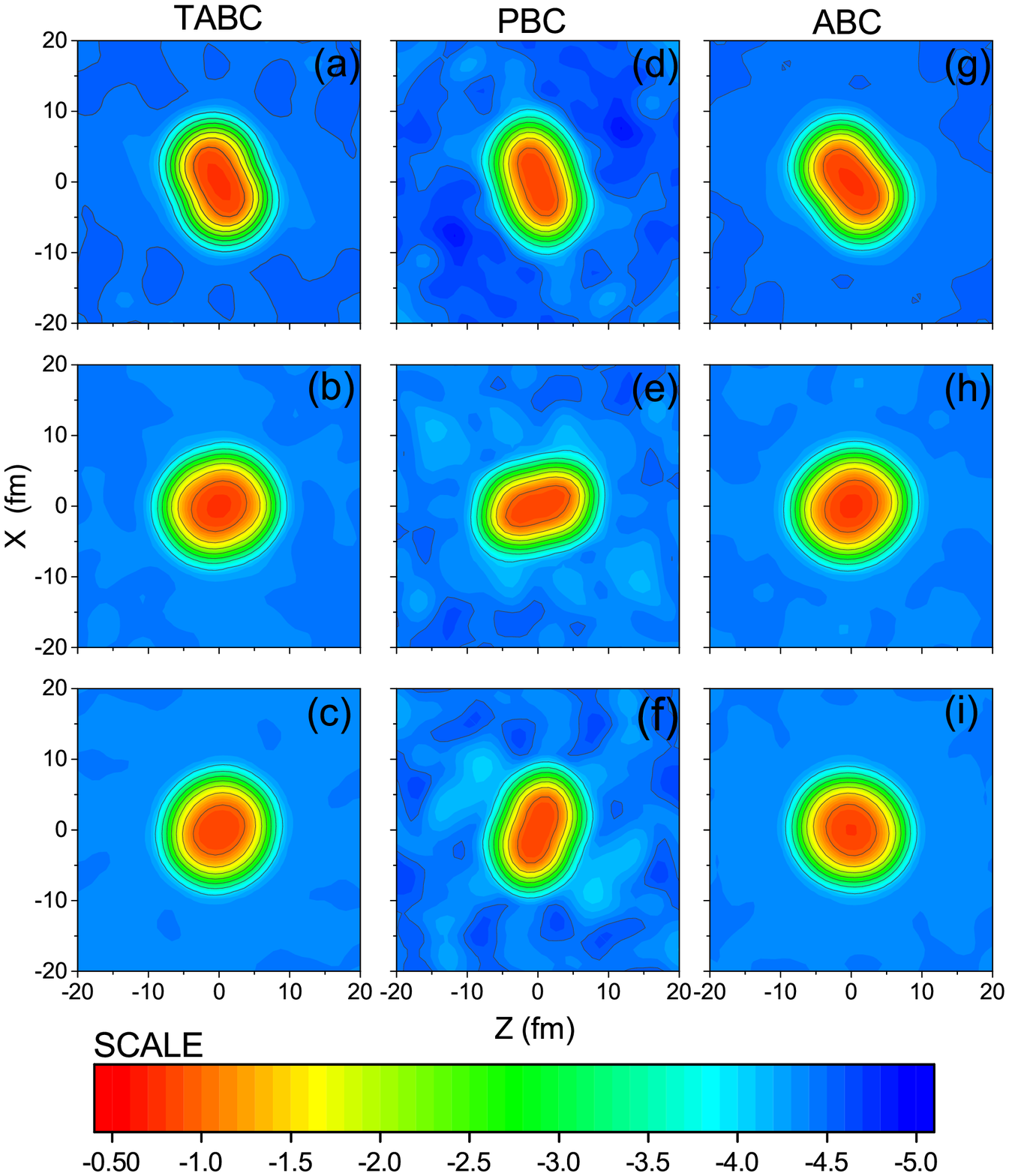}\\
  \caption{(Color online)  The evolution of density distributions (in x-z plane) in log scale, corresponding to Fig.\ref{fig5}. TABC results are shown on the left as (a)t=2000 fm/c; (b)4000 fm/c; (c) 6000 fm/c. PBC results are shown in the middle as (d) t=2000 fm/c; (e) 4000 fm/c; (f) 6000 fm/c. ABC results are shown on the right as (g)t=2000 fm/c; (h)4000 fm/c; (i) 6000 fm/c. }
  \label{fig6}
\end{figure}

Fig.\ref{fig1} shows the time evolution of the total kinetic energy for head-to-head collisions of $^{24}$Mg+ $^{24}$Mg, with collision energies at $E_{c.m}$=29, 49 and 69 MeV, respectively.
The corresponding excitation energies of the compound $^{48}$Cr are 43.93, 63.93, 83.93 MeV.
In the fusion stage, there are strongly damped oscillations in kinetic energies related to the bulk dissipation.
In this stage, there are negligible differences between PBC and TABC calculations before $t<1000$ fm/c.
In PBC calculations,  small amplitude oscillations are persistent after $2000$ fm/c.
These small amplitude oscillations behave like molecular vibrating states and are presented at three different collision energies.
In TABC calculations, however, the small amplitude oscillations are quickly damped and the compound nucleus at equilibrium is obtained.
This damping effect has been demonstrated in the TDHF calculations of strengthes of giant resonances with TABC and ABC~\cite{schuer}.
The particles are not really escaped in TABC,  however, the box-size dependence of continuum treatment is actually diminished in TABC.
For PBC calculations with not sufficiently large box sizes, the continuum is not precisely discretized.
 The continuum damping plays a persistent role after the fusion stage.
This means that ``molecular vibration states" obtained in TDHF calculations with PBC are questionable.

Fig.\ref{fig2} displays the Fourier analysis of the evolution of kinetic energies from TDHF-TABC calculations as shown in Fig.\ref{fig1}.
It can be seen that for the three collisions at energies of 29, 49 and 69 MeV, the main peaks are at 5.8, 5.0  and 6.0 MeV, which are lower than that of typical giant resonances~\cite{pring}.
The damping widths for $E_{c.m}$=29, 49, 69 MeV are about 2.2 MeV, 1.4 MeV and 2.3 MeV, respectively.
The head-to-head fusion is a typical underdamping process with damping widths much smaller than oscillation frequencies, in contrast to the widely recognized overdamped fission process~\cite{overdamp,overdamp2}.
The main dampings are more or less similar in three cases. It can be seen that the damping time of the 49 MeV collision is longer than other two cases, which is related
to its narrower damping width.
For PBC calculations, the small amplitude oscillations correspond to frequencies of 7.4, 7.0, 7.6 MeV, respectively. These frequencies are much smaller than the Ikeda clustering threshold energies and are not
likely to be physical molecular vibrations.

The time evolutions of quadrupole deformations from TDHF-TABC calculations of the above-mentioned collisions are shown in Fig.\ref{fig3}.
At the fusion stage, the large-amplitude oscillations are strongly damped.
It can be seen that the amplitudes are larger with higher collision energies.
However, the minimum deformations ($\approx$260 fm$^{2}$) in the three fusion processes are close.
The oscillations of deformations are not as symmetric as that in kinetic energies.
The fusion is not a simple damped oscillator regarding the quadrupole deformations, due to
the density dependence of incompressibility.
At the equilibrium stage, the quadrupole deformations are 448, 485, and 510 fm$^{2}$, respectively.
These prolate deformations of the compound $^{48}$Cr are extremely large with axis ratios are 3.9:1, 3.8:1, and 3.2:1, respectively.
Therefore larger quadrupole deformations at higher collision energies are not necessarily related to larger axis ratios, and volume expansions of compound nuclei  play a role.
The corresponding final kinetic energies of the three cases in Fig.\ref{fig1} are 783.5, 776, and 768 MeV respectively, which
are lower than the initial kinetic energies at 809, 829, and 849 MeV respectively. The differences
indicate that the total potential energies increase significantly as deformation energies play a role.

Fig.\ref{fig4} shows the time evolutions of the $^{24}$Mg+ $^{24}$Mg side-to-side collision at 49 MeV.
Similar to Fig.\ref{fig1}, the PBC and TABC calculations are close at the fusion stage before 1000 fm/c.
This process is complex and doesn't like a damped oscillator as in the head-to-head fusion.
Small amplitude oscillations in quadrupole deformations and kinetic energies
are also persistent in PBC calculations, but they are being damped in TABC calculations.
This is the same continuum damping as demonstrated in the head-to-head collisions.
The final kinetic energy is about 824 MeV, which is larger than 776 MeV of the head-to-head collision with the same collision energy.
The final quadrupole deformation is about 220 fm$^{2}$ which is much smaller than that of the head-to-head collision, showing the role of deformation energy and dependence of collision orientations.

The last but most interesting part of this work is the fusion-rotation reaction. In this case, the collision energy is taken as 40 MeV and the impact parameter is 2 fm for the side-to-side collision.
In our calculations, the collision direction is along z-axis and the compound nucleus rotates in the x-z plane.
To study the rotation evolution, the expectation values of $<z^2>$ are given in Fig.\ref{fig5}(a).
It is striking to see that the rotation amplitudes are slowly damped in  TABC and ABC calculations, while the rotation is almost a perfect
cosine function in PBC calculations.
This rotational damping is not a surprise considering the pervious damping of small amplitude vibrations.
It is more interesting because it illustrated a very clear damping picture compared to previous vibrational cases. The vibrational damping has been studied extensively~\cite{bertsch}, while
the rotational damping has rarely been discussed~\cite{rotdamp}.
To further study the damping effects, the spectral analysis by Fourier transformation are also shown in Fig.5(b).
Note that the $<z^2>$ frequency (period is about 1300 fm/c) is two times the rotation frequency.
The resulted rotation frequencies $\omega$ are 0.43, 0.46 and 0.49 MeV for TABC, PBC and ABC calculations, respectively.
ABC calculations resulted in a larger rotational frequency (in Fig.\ref{fig5}(a)) and angular momentum (in Fig.\ref{fig5}(c)).
Note that the damped rotation amplitudes can be written as $A(t)=A_0e^{-\Gamma t/2}\cos(\omega_Dt+\phi) $,
where the damped frequency $\omega_D= \sqrt{\omega^2-\Gamma^2/4}$.
Therefore $\omega_D$ of TABC is slightly smaller than that of PBC due to damping effects.
The damping width $\Gamma$  can be estimated to
be about 0.16 MeV for TABC and ABC calculations, while $\Gamma$ is about 0.08 MeV for PBC calculations.
The obtained damping widths are comparable with results from cranking shell model calculations of compound nuclei~\cite{rotdamp},
which can be measured by experiments~\cite{rotdamp2,rotdamp3}.

Fig.\ref{fig6} shows the evolutions of density distributions in the x-z plane, corresponding to rotations in Fig.5.
With time increases, we see that the surface density distributions becomes more and more uniform in TABC and ABC calculations.
Note that surface gases in TDHF calculations are dynamic rather than static.
The gases in ABC calculations are being absorbed at outer boundaries as
well as being produced simultaneously. Consequently, the particle numbers with ABC are not conserved. 
The gas density is about $10^{-5}$ fm$^{-3}$, to which protons and neutrons have similar contributions and $\alpha$ emission is possible.
The uniform gas is similar to the thermal gas (dominated by neutrons) in heavy compound nuclei from finite temperature Hartree-Fock-Bogoliubov calculations~\cite{zhu}.
The thermal neutron escaping width is proportional to the gas density which provides an equilibrium pressure.
The density of thermal neutron gas is dependent on the temperature and independent of box sizes.
The surface gas corresponds to emitted particles and in principle it should be removed.
We see PBC, TABC and ABC all lead to a low-energy bump at $\hbar\omega$=0.2 MeV in Fig.\ref{fig5}(b), due to the floating continuum gases, as discussed in \cite{schuer}.
In contrast, the density distributions at surfaces in PBC calculations have non-uniform structures.
The non-uniform surface density is related to  finite box-sizes, which  doesn't cause damping effects in Fig.\ref{fig5}.
The quadrupole deformation of PBC calculations is about 227 fm$^2$ with an axis ratio of 2.2:1.
In TABC and ABC calculations, however, the averaged bulk density distributions evolve towards spherical.
This is consistent with the damped rotation amplitude in Fig.\ref{fig5}.
In head-to-head collisions, differently, large equilibrium deformations are maintained in compound nuclei.

TABC calculations show that the near-spherical compound nucleus $^{48}$Cr is still rotating at each twisted angles. For the rotation in x-z plane,
 the calculated averaged angular momentum  $<$$J_y$$>$ is about 6$\hbar$ at 7000 fm/c, which decays slowly and smoothly.
The calculated $<$$J_y$$>$ with PBC is decaying slightly faster associated with oscillations.
 It is known that hot nuclei at equilibrium become spherical at high temperatures in the mean-field framework.
 In our case, the angular momentum $<$$J_y$$>$ from TABC is persistent although density distributions become spherical.
It is understandable that spherical compound nuclei at equilibrium can rotate due to the fading of quantum effects, although spherical quantum systems don't rotate.
In Fig.\ref{fig5}(c), we see that the total angular momentum is not conserved with different boundary conditions.
In TABC and PBC, the angular momentum is not conserved whenever the emitted particles from highly-excited compound nuclei encounter boundaries, while
this conservation is preserved for TDHF cranking calculations~\cite{guo}. In this respect, ABC is more reasonable since its angular momentum  $<$$J_y$$>$ decreases slowly
due to real particle emissions and $<$$J_y$$>$ is larger than TABC results.

\emph{Summary.}---
We implemented the twisted boundary condition in time-dependent Skyrme Hartree-Fock calculations of $^{24}$Mg +$^{24}$Mg collisions,
which are performed in 3D coordinate spaces.
In head-to-head and side-to-side collisions, small amplitude vibrations are persistent in calculations with periodic boundary conditions,
but they are damped with twist-averaged and absorbing boundary conditions.
In TABC, this kind of damping mechanism is related to the cancelation of box size dependence in continuum treatment.
By studying the side-to-side collision with an impact parameter of 2 fm, we found that the rotation amplitude is damped as well, in which
the continuum damping width can be clearly extracted.
The density distributions show that in TABC and ABC calculations the compound nucleus becomes spherical surrounded by a uniform gas.
The surface density distributions in PBC calculations are non-uniform, due to finite box-size effects.
The angular momentum decreases slowly due to particle emissions.
The near-spherical compound nucleus are still rotating with a considerable angular momentum.
These results are inspiring and provide a better understanding of rotating compound nuclei.
In principle, ABC can be applied to a finite system with continuum by avoiding wave reflections.
PBC descriptions are insufficient for a system when continuum is not
negligible. TABC can remedy the spurious effects 
due to finite box sizes
by describing the periodicity of a
system with continuum correctly.  With a very large box, different boundary conditions should give consistent results~\cite{abc1}. 
We demonstrated that consequences of continuum damping after long-time evolutions could be significant.
Further applications of twisted boundary conditions in nuclear reactions and weakly bound nuclei will be
valuable.

\begin{acknowledgments}
We are grateful to W. Nazarewicz 's suggestions and useful discussions on twisted boundary conditions.
We also thank useful discussions with F.R.Xu and P. Stevenson.
 This work was supported by  National Key R$\&$D Program of China (Contract No. 2018YFA0404403),
 and the National Natural Science Foundation of China under Grants No.11790325,11522538,11835001.
We also acknowledge that computations in this work were performed in Tianhe-1A
located in Tianjin and Tianhe-2
located in Guangzhou.
\end{acknowledgments}

\nocite{*}


\begin{thebibliography}{999}

\bibitem{negele}
J. W. Negele, Rev. Mod. Phys., 54, 9135(1982).

\bibitem{tddft}
T. Nakatsukasa, K. Matsuyanagi, M.Matsuo, and K. Yabana,
Rev. Mod. Phys. 88, 045004(2016).

\bibitem{tdhf}
C. Simenel, Eur. Phys. J. A  48, 152(2012).

\bibitem{umar}
A.S. Umar, V.E. Oberacker, Nucl. Phys. A 944, 238(2015).

\bibitem{guo}
L. Guo, J. A. Maruhn, P.-G. Reinhard, Y. Hashimoto, Phys. Rev. C 77,041301(R)(2008).

\bibitem{scamps}
G. Scamps and D. Lacroix, Phys. Rev. C 87, 014605 (2013).

\bibitem{paul}
P.Goddard, P.Stevenson, and A. Rios,
Phys. Rev. C 92, 054610 (2015).

\bibitem{tdhfb1}
A.Bulgac, P.Magierski, K.J. Roche, and I.Stetcu,
Phys. Rev. Lett. 116, 122504 (2016).

\bibitem{tdhfb2}
P.Magierski, K. Sekizawa, and G. Wlazlowski,
Phys. Rev. Lett. 119, 042501 (2017)

\bibitem{qmd}
J. Aichelin, Phys. Rept. 202, 233(1991).

\bibitem{umar05}
A.S. Umar and V.E. Oberacker, Phys. Rev.C 71, 034314(2005).

\bibitem{sky3d}
J. A. Maruhn, P.-G. Reinhard, P. D. Stevenson, and A. S. Umar,
Comp. Phys. Comm. 185, 2195 (2014).


\bibitem{abc1}
P.-G. Reinhard, P. D. Stevenson, D. Almehed, J. A. Maruhn, and
M. R. Strayer, Phys. Rev. E 73, 036709 (2006).

\bibitem{abc2}
T.Nakatsukasa, K.Yabana, Phys. Rev. C 71, 024301(2005).

\bibitem{cdcc}
M. Yahiro, K. Ogata, T. Matsumoto, K. Minomo, Prog. Theor. Exp. Phys. 2012, 01A206(2012).

\bibitem{zhu}
Y. Zhu, and J. C. Pei,
Phys. Rev. C 90, 054316 (2014).

\bibitem{matsuo}
M. Matsuo,  Nucl. Phys. A 696, 371 (2001).

\bibitem{zhang}
J.C. Pei, Y.N. Zhang, F.R. Xu, Phys. Rev. C 87, 051302(R)(2013);
Y.N. Zhang, J.C. Pei, and F.R. Xu, Phys. Rev. C 88, 054305(2013).


\bibitem{sun}
Z.H. Sun, Q. Wu, Z.H. Zhao, B.S. Hu, S.J. Dai, and F.R. Xu,
Phys. Lett. B 769 227 (2017).


\bibitem{schuer}
B. Schuetrumpf, W. Nazarewicz and P.-G. Reinhard, Phys. Rev. C, 93, 054304 (2016).

\bibitem{lin}
C. Lin, F. H. Zong, and D. M. Ceperley, Phys. Rev. E, 64, 016702(2001).

\bibitem{luu}
C.Korber and T.Luu, Phys. Rev. C 93, 054002(2016).

\bibitem{wang}
K. Wang, M. Kortelainen, and J. C. Pei, Phys. Rev. C 96, 031301(R) (2017).

\bibitem{high-spin}
T. Inakura, S. Mizutori, M. Yamagami, K. Matsuyanagi, Nucl. Phys. A 710, 261(2002).


\bibitem{nitto}
A. Di Nitto, et al., Phys. Rev. C 93, 044602(2016).

\bibitem{wuossma}
A. H. Wuosmaa et al., Phys. Rev. Lett. 58, 1312 (1987).

\bibitem{zurmuhle}
R. W. Zurmuhle et al. , Phys. Lett. 129B, 384 (1983).

\bibitem{Jachcinski}
C. M. Jachcinski, D. G. Kovar, R. R. Betts, C. N. Davids, D. F. Geesaman, C. Olmer, M. Paul, S. J. Sanders, and J. L. Yntema,
Phys. Rev. C 24, 2070(1981).

\bibitem{zhou}
B. Zhou, Y. Funaki, H.Horiuchi, Z.Z. Ren, G. Ropke, P. Schuck, A. Tohsaki, C. Xu, and T. Yamada,
Phys. Rev. C 89, 034319 (2014).

\bibitem{schuer3}
B. Schuetrumpf and W. Nazarewicz,
Phys. Rev. C 96, 064608(2017).

\bibitem{svbas}
P. Klupfel, P.-G. Reinhard, T. J. Burvenich, and J. A.
Maruhn, Phys. Rev. C 79, 034310 (2009)

\bibitem{dai}
G.F. Dai, L. Guo, E.G. Zhao, S.G. Zhou,Phys. Rev. C 90, 044609 (2014).

\bibitem{schuer2}
B. Schuetrumpf, W.Nazarewicz, Phys. Rev. C 92, 045806(2015).

\bibitem{bender}
M. Bender, P.-H. Heenen, and P.-G. Reinhard,
Rev. Mod. Phys. 75, 121 (2003).

\bibitem{wangm}
M. Wang, G. Audi,F.G. Kondev, W.J. Huang,
S. Naimi, X. Xu, Chin. Phys. C 41, 030003(2017).

\bibitem{pring}
G.F. Bertsch and R.A. Broglia, \textit{Oscillations in Finite Quantum Systems}, (Cambridge University Press, 1994).

\bibitem{overdamp}
 A. E. Gettinger and I. I. Gontchar, J. Phys. G 26, 347(2000).

\bibitem{overdamp2}
J. Randrup and P. Moller, Phys. Rev. C 88, 064606(2013).

\bibitem{bertsch}
G.F. Bertsch, P.F. Bortignon, R.A. Broglia, Rev. Mod. Phys. 55, 287(1983).

\bibitem{rotdamp}
M. Matsuo, S. Leoni, C. Grassi, E. Vigezzi, A. Bracco, T. Dossing, B. Herskind, AIP Conf.Proc. 656, 32(2003).

\bibitem{rotdamp2}
F.S. Stephens, M.A. Deleplanque, I.Y. Lee, D. Ward, P. Fallon, M. Cromaz, R. M. Clark, R. M. Diamond, A. O. Macchiavelli, and K. Vetter,
Phys. Rev. Lett. 88, 142501(2002).

\bibitem{rotdamp3}
S. Leoni et al.,
Phys. Rev. Lett. 93, 022501(2004).


\end{thebibliography}

\end{document}